\documentclass[review]{elsarticle}

\usepackage{subcaption}
\usepackage{hyperref}
\usepackage{multicol}
\usepackage{colortbl}
\usepackage{tabularx}
\usepackage{multirow}
\usepackage{booktabs}
\usepackage{amsmath}
\usepackage{rotating}

\journal{Energy Policy}

\biboptions{authoryear}

\begin{document}

\begin{frontmatter}

\title{Slipping through the net: Can data science approaches help target clean cooking policy interventions?}

\author[mymainaddress]{Andr\'e Paul Neto-Bradley\corref{mycorrespondingauthor}}
\cortext[mycorrespondingauthor]{Corresponding author}
\address[mymainaddress]{Department of Engineering, University of Cambridge, Cambridge, UK}
\address[mymainaddress0]{Data Centric Engineering, Alan Turing Institute, London, UK}
\address[mymainaddress1]{Indian Institute for Human Settlements, Bangalore, India}
\ead{apn30@cam.ac.uk}

\author[mymainaddress,mymainaddress0]{Ruchi Choudhary}
\author[mymainaddress1]{Amir Bazaz}

\begin{abstract}
Reliance on solid biomass cooking fuels in India has negative health and socio-economic consequences for households, yet policies aimed at promoting uptake of LPG for cooking have not always been effective at promoting sustained transition to cleaner cooking amongst intended beneficiaries. This paper uses a two step approach combining predictive and descriptive analyses of the IHDS panel dataset to identify different groups of households that switched stove between 2004/5 and 2011/12. A tree-based ensemble machine learning predictive analysis identifies key determinants of a switch from biomass to non-biomass stoves. A descriptive clustering analysis is used to identify groups of stove-switching households that follow different transition pathways. There are three key findings of this study: firstly non-income determinants of stove switching do not have a linear effect on stove switching, in particular variables on time of use and appliance ownership which offer a proxy for household energy practices; secondly location specific factors including region, infrastructure availability, and dwelling quality are found to be key determinants and as a result policies must be tailored to take into account local variations; thirdly some groups of households that adopt non-biomass stoves continue using biomass and interventions should be targeted to reduce their biomass use.

\section*{Highlights}
\begin{itemize}
   \item Policies promoting cleaner cooking do not reach all intended beneficiaries.
   \item Descriptive analytics identify household groups with distinct transition pathways.
   \item Non-income factors do not all linearly affect the probability of switching stove.
   \item Policies must be tailored to take into account local socio-economic variations. 
   \item Clean cooking interventions must address residual biomass use.
\end{itemize}
\end{abstract}
\begin{keyword}
Energy Access, Cooking Fuel, Energy Poverty, India, Urban Analytics
\end{keyword}
\end{frontmatter}

\section{Introduction}
Worldwide there are almost 3 billion people who do not have access to clean cooking fuel, and in India just under half the population still face limited access to clean cooking fuels \citep{international_energy_agency_main_2019}. Reliance on solid fuels has negative consequences including the health impacts of household air pollution, environmental impacts of local deforestation, and negative socio-economic effects arising from the practices surrounding the use of such biomass fuels \citep{smith_making_2014}. These socio-economic effects disproportionately impact women and children of the household, for example the time spent collecting fuel by female members of the household negatively impacts their livelihoods and empowerment \citep{rahut_patterns_2016}. While in the past there were attempts to improve biomass stove efficiency to reduce negative health impacts of air pollution from biomass fuel use, there has been growing recognition that solving the wider negative socio-economic and health impact of solid fuel use requires a transition towards cleaner alternatives such as gas and electricity \citep{batchelor_two_2019}.

The residential sector in India accounts for 29\% of the country's total energy consumption, and is second only to the industrial sector \citep{international_energy_agency_india_2020}. However commercial sources of energy in the form of electricity and petroleum products account for only around 30\% of household energy consumption, although this share is set to continue growing \citep{ministry_of_statistics_and_programme_implementation_energy_2019}. 68\% of energy consumed by households comes from biomass and waste, which is primarily used for cooking and heating and accounts for almost two thirds of all biomass use in India \citep{international_energy_agency_india_2020}.

In recent years there has been a concerted effort in India to promote the uptake of Liquified Petroleum Gas (LPG) for cooking to reduce the use of solid fuels and tackle the associated negative health and development consequences of their use. Most recently the flagship Pradhan Mantri Ujjwala Yojana (PMUY) programme achieved its target of providing 80 million low-income households with a gas connection. The PMUY programme provided financial support to below poverty line households, covering the cost of LPG connection and subsidising the first LPG cylinder (amounting to around 60\% of total initial cost of 5000 INR), thus reducing the initial cost barrier \citep{sharma_transition_2019}. However, studies have found that while the programme successfully enabled many households to acquire their first cylinder, several of those households have not necessarily transitioned to sustained LPG use. They continue to use solid fuels for part or all of their needs \citep{kar_using_2019}. Further to that, findings suggest that the programme may not have managed to reach its intended beneficiaries equally in all regions, and benefited some households that would likely have transitioned without the incentive from the programme \citep{sharma_transition_2019, sankhyayan_availability_2019}. These are outcomes also seen in other top-down clean energy interventions in India and the Global South more widely \citep{sehjpal_going_2014, silver_powering_2017, kebede_can_2002}. Whilst many energy poor households may benefit from such policies, there are always those who don't benefit as expected, or 'slip through the net' and miss out altogether \citep{rao_kerosene_2012,batchelor_two_2019}.

PMUY is the most recent in a long line of policies aimed at reducing costs of LPG for poor households and improving access. In the last two decades there have been a range of other programmes and initiatives to promote the uptake of LPG including the 'Vitrak Yojana' from 2009 which aimed to increase LPG distributorship through improving infrastructure and supply chains, while a range of subsidies have existed at national and local levels \citep{sankhyayan_availability_2019}. Recently there has been an initiative in place alongside PMUY to encourage wealthier households to voluntarily give up their subsidy if they do not need it so that it may benefit a poorer household \citep{sharma_transition_2019}.

A key assumption underpinning these policies is that use of a cleaner fuel, in this case LPG, is desired by all households and that the barrier that prevents them from using this fuel is the upfront cost of switching \citep{kar_using_2019}. This simplifies lack of access to clean cooking to an issue of household income or lack thereof, understating the complexity of barriers to clean cooking transitions \citep{sankhyayan_availability_2019}. As \cite{gould_lpg_2018} show, for many rural households in India the upfront cost is only one of many barriers to sustained LPG use, with other notable barriers including the lump-sum nature of monthly payments for LPG cylinders (as opposed to the actual cost) as well as the time and difficulty of transporting the cylinders.

Lack of access to clean cooking can be a form of energy poverty, and as described by \cite{sadath_assessing_2017} problems of energy poverty are multidimensional and should not be simply confused with income poverty. \cite{khandker_are_2012} showed that income non-poor households were not necessarily energy non-poor, and the effect of non-income variables on energy decision making plays a key role in determining the energy poverty of a household. There is a substantial body of literature using quantitative methods to model household energy use. Predictive modelling efforts have shown the relationship between appliance ownership and electricity demand \citep{murthy_end-uses_2001, van_ruijven_model_2011}, and studies have identified socio-economic determinants of appliance ownership including education, caste, and dwelling characteristics \citep{rao_white_2017,dhanaraj_income_2018}. Studies have shown of a hierarchy of preference of fuels in India \citep{farsi_fuel_2007,ekholm_determinants_2010}, and have used income as the basis for modelling consumption of clean fuels \citep{pachauri_measuring_2004, filippini_elasticities_2004, lam_kerosene_2016, bhattacharyya_influence_2015}. Quantitative modelling has also identified non-income socio-economic determinants of clean fuel use including profession of the head of the household \citep{kemmler_factors_2007, sehjpal_going_2014}, education \citep{ahmad_fuel_2015, sankhyayan_availability_2019}, and land ownership \citep{sehjpal_going_2014}, although findings across all these studies show considerable variation \citep{chunekar_towards_2019}.

A macro scale technical-economic view of energy transition which focuses on cost and performance of different alternatives and assumes that households behave as rational consumers has proven useful for quantitatively understanding energy consumption trends. However, \cite{van_der_kroon_energy_2013} make the case that identifying different energy transition pathways requires a better understanding of the decision-making and external context of a household. This requires understanding the social aspects of energy use, through preferences, practices, and decisions of households which act at a local scale. 

Social practice theory (SPT) provides a lens through which energy transitions at a local scale can be analysed. SPT approaches used in energy research follow the approach of \cite{shove_dynamics_2019} which views people as 'practitioners' who combine materials, competences or know-how, and meanings to create practices \citep{bisaga_climb_2018, khalid_homely_2017}. This provides a shift in perspective away from macro scale resourced-based systems thinking towards more micro scale individual enquiry of what energy is actually used for.

Several recent studies have applied an SPT approach to the study of energy use in the Global South. In their work on middle class households in Pakistan, \cite{khalid_homely_2017} found that household practices shaped around cultural norms and socio-cultural dynamics explained the peculiar nature of electricity demand of these households. Similarly \cite{bisaga_climb_2018} found that understanding the practices surrounding lighting and mobile phone charging helped to explain energy demand and transition in Rwandan rural-off grid communities. A more recent study by \cite{debnath_how_2019} provides an Indian context using a quantitative approach grounded in SPT to explore the influence of non-income factors on appliance ownership in rehabilitated slums in Mumbai, finding that the change in built environment following the rehabilitation of slums led to a change in practices which translated into changes of appliance ownership and usage. However the population samples in all of these studies were relatively homogeneous in their socio-economic characteristics, and differences in practices are observed on a household to household level. \cite{galvin_schatzkian_2016} explain that socio-economic causality in energy consumption studies can be a blind spot for SPT approaches, and households in such studies are often relatively uniform in their socio-economic profiles.

Findings at a macro-scale may point to the existence and prevalence of certain trends governing uptake of modern fuel use. For example, \cite{rao_white_2017} found that Sikh households in India were more likely to own a refrigerator. However, correct interpretation of the causes and implications of such trends sometimes requires taking account of local practices, behaviours, and decisions that define energy use. For the refrigerator ownership trend above knowledge of practices and behaviours, in this case the higher consumption of dairy products by Sikhs compared to other religious groups in India, offered an explanation for this trend.

There has been growing interest in the role of urban data analytics for improving our understanding of energy provision \citep{bibri_smart_2017}, and these are particularly relevant given the multi-dimensional nature of energy poverty. Such studies use techniques from the data sciences to process large socio-economic and/or demographic datasets with mixed datatypes to inform better policy interventions. This involves three broad categories of analysis: descriptive analysis, which is concerned with understanding the data; predictive analysis which is concerned with extrapolating the trends found in the data; and prescriptive analysis which is concerned with using the data to identify the interventions likely to achieve desired outcomes \citep{wang_review_2019}. The majority of macro scale quantitative studies make use of regression models, which constitute a form of predictive analysis. However, prediction models based on regression assume that all variables considered in the analysis are independent and influence a given quantity of interest in a similar manner. This restricts the understanding of variations of features that influence uptake of clean fuels across different types of households and their local context. Descriptive analysis overcomes these limitations, and can enhance the understanding of the composition of characteristics that govern energy transitions. Together with predictive analysis, it can yield a better understanding of the different sets of features that act as barriers to transition between groups of households so as to support targeted policy interventions. While previous studies have made use of novel data science techniques to perform predictive regression analyses such as \cite{rao_white_2017}, the authors are not aware of studies combining these with descriptive data science methods to support targeted policy interventions. In this paper, we demonstrate a novel combination of descriptive and predictive analysis to support the design of more targeted clean cooking policies. 

We employ ensemble machine learning and clustering algorithms to conduct a combined predictive and descriptive analysis of the panel data from the Indian Human Development Survey (IHDS) between 2004/5 and 2011/12. The predictive analysis characterises the non-linear relationship of different determinants that influence clean cooking adoption. Through the descriptive analysis, we demonstrate that groups of households follow different cooking transition pathways. The combined analysis identifies differences in key policy features between groups of households for which targeted interventions could be designed to address the needs and challenges of households that might currently be under-served by cost-centric policies. The remainder of this paper is structured as follows: section 2 discusses the features and handling of the dataset, section 3 describes our analytical methods. Section 4 presents results of two different regression models and compares the performance of these, leading on to section 5 which presents the results of a clustering analysis of households that did switch to a non-biomass stove discussing the existence of different types of switching household, discussing limitations in section 6 and presenting conclusions and policy implications in section 7.

\section{Data}
Our study uses household level survey data from the publicly available and nationally representative IHDS. The first IHDS was conducted in 2004-2005 (referred to as IHDS-I) \citep{desai_india_2010} with a second follow-up survey in 2011-2012 (IHDS-II) \citep{desai_india_2015}, which returned to survey the same households originally surveyed for IHDS-I. The surveys were conducted by means of two one-hour interviews with the whole household or the head of the household, and comprised of a nationally representative sample of 41,554 urban and rural households across all Indian territories excluding the Andaman Isles and Lakshadweep. This sample included 1503 villages and 971 urban city blocks across 383 districts in 33 different states. IHDS-II covered 85 percent of the original households, with those households not surveyed the second time either having been unreachable, having moved, or been struck by natural disaster \citep{desai_india_2015}. To control for the effect of changes in built environment and constitution of the household due to a move or split of the household we included only instances where the original household was recontacted. Thus the 6,911 households from the IHDS-I that could not be recontacted were excluded as well as the 1,721 households which had split into multiple different households. This analysis uses the resulting subset of 32,922 households which were surveyed both in the IHDS-I and IHDS-II.

As pointed out by \cite{khandker_are_2012}  and \cite{ahmad_fuel_2015}, the energy related questions in the IHDS are more comprehensive than those in comparable studies including the Living Standards Measurement Studies coordinated by the World Bank, and the NSS Surveys of Consumer Expenditure. The IHDS dataset is disaggregated by housing type and various demographic features such as gender, religion, caste, occupation, and education \citep{desai_india_2015,desai_india_2010}. Additionally the IHDS includes some information on time spent carrying out certain energy related practices in the household, including time spent watching television, time spent collecting firewood, and hours of stove usage. Recommendations from the authors of the dataset were followed \citep{desai_india_2015} with regards to weightings and variable selection. All weightings used were the 'SWeights' specified for the households in the IHDS-I, and values for relatively unchanging variables (e.g. Caste and Religion) were taken from the IHDS-II.

The main dependant variable of interest was a binary variable indicating whether the household had switched from primarily using a biomass stove in 2004/5 to a non-biomass stove in 2011/12. This was constructed using the the variables indicating the main stove used for cooking in the IHDS-I and IHDS-II respectively. The stove options included 3 types of biomass solid fuel stoves, and a general 'modern stove' category which could represent kerosene, LPG, or electric stoves. In our IHDS dataset 5358 households switched from using a biomass stove as their primary stove in 2004-5 to using a 'modern' non-biomass stove in 2011-12 representing 16.27\% of households (14.94\% when adjusted by sampling weights). 

Other variables were constructed from the dataset either to make variables more comparable or to create a dummy variable for a particular characteristic, or to characterise change in a variable between surveys. Energy consumption values in the IHDS are given in units of cost (INR) as opposed to units of energy which makes comparisons difficult. These values were converted to estimated energy consumption in kWh using local price data available in the IHDS and collected from government sources \citep{government_of_india_planning_commission_annual_2012}. In addition appliance ownership was grouped according to associated household activity: cooking (Pressure Cooker, Mixer/Grinder, Microwave, Refrigerator), and IT (Television, Telephone, Mobile Telephone, Computer, Laptop).

\section{Methods}
Studies on energy transition typically use some form of logit or probit regression model to perform a predictive analysis identifying the trends and effect of a given set of variables on appliance ownership, fuel use, or adoption rates of electricity or LPG. Recently \cite{rao_white_2017} used a form of ensemble technique called a Boosted Regression Tree (BRT) model to analyse the effect of a range of household characteristics on the uptake of so-called 'white good' appliances. A comparison of the predictive capability of these two modelling approaches found that the BRT model on the whole outperformed the logit model in predicting appliance ownership \citep{rao_white_2017}.

In this study we seek to provide a greater level of descriptive or explanatory analysis to identify the different transition pathways and a two stage approach was used to achieve this. The first stage involves predictive modelling using an ensemble machine learning technique to identify factors that are determinants of clean cooking transition and assess performance of the model. The second stage focuses on descriptive modelling using hierarchical clustering, where the key determinants identified in the predictive modelling are used to identify the different groups of households that did switch stove and the different combination of features that characterize each group. The first stage of the analysis uses a training subset of 25,000 of the 32,922 households from the IHDS to identify the influence of variables on the propensity of a household to switch from a solid fuel biomass stove to a cleaner 'modern stove' as their main cooking stove. The predictive performance of the ensemble learning regression and a conventional probit regression are assessed and compared using the remainder of the dataset not used to train the model. The secondary stage of analysis uses agglomerative hierarchical clustering to cluster the 5,358 households that did switch from biomass to a 'modern' non-biomass stove. By comparing the effect of key determinants identified by the predictive modelling and the defining characteristics of the clusters of stove-switching households, it is possible to identify the different combinations of key determinants enabling stove transition and policy features in each cluster. 

Variable selection was carried out using both correlation and random forest analysis to identify the most relevant variables. Given the inter-related nature of the socio-economic and cultural variables of interest in the dataset it was important to identify and address any significant multi-collinearity in the dataset before performing any analysis. A Farrar-Glauber test was conducted to identify and address any multi-collinearity. In particular fuels used exclusively for cooking showed cross-dependent correlation with one another, so redundant fuels were removed from the selected variables. In addition, the number of different region categories was reduced by reassigning households in states in the center region to the neighbouring eastern region as there was little distinction between these two. The descriptive statistics of the resulting independent variables are show in table \ref{Table of Variables} (except profession, caste, and region which are non-continuous, and non-binary).

\begin{table}[h!] \centering 
  \caption{Descriptive statistics for variables} 
  \label{Table of Variables} 
\begin{tabular}{lcccc} 
\\[-5ex]\hline 
\hline \\[-4ex] 
Independent variable & Mean & Median &  Min. &  Max.  \\ 
\hline \\[-4ex] 
 Income per capita (INR/month)  & 2401 & 1363 & 0 & 346750  \\ 
 Urban & 0.330 & - & 0 & 1 \\
 Time in Place (years) & 78.41 & 90.00 & 0.00 & 90.00  \\ 
 Female Education (years) & 5.395 & 5.000 & 0.000 & 16.000 \\
 Permanent House & 0.704 & - & 0 & 1 \\ 
 Flush Toilet & 0.392 & - & 0 & 1  \\ 
 Piped Water Availability (hours/day) & 1.845 & 0.000 & 0.000 & 24.000 \\ 
 Dairy Spend (INR/month) & 196.40 & 100.00 & 0.00 & 8600 \\ 
 Electricity Availability (hours/day) & 13.11 & 14.00 & 0.00 & 24.00 \\ 
 Electricity Consumption (kWh/month) & 93.68 & 54.50 & 0.00 & 1977.40 \\ 
 Kerosene Consumption (kWh/month) & 28.01 & 24.79 & 0.00 & 587.00 \\ 
 Change in fuel collection time (min) & -3.475 & 0.000 & -320.000 & 450.000  \\ 
 Cooking appliance ownership & 0.291 & 0.250 & 0.000 & 1.000  \\ 
 IT appliance ownership & 0.310 & 0.429 & 0.000 & 1.000 \\ 
 Change in Female TV Time (hours/day) & 0.687 & 1.000 & -12.000 & 14.000 \\ 
\hline \\[-4ex] 
\end{tabular} 
\end{table} 

The BRT is a tree based ensemble learning technique that combines a large number of simple categorisation trees, using gradient boosting to build ensembles of decision trees that are fit to the remaining model residuals. Unlike a probit model there is no a priori specification of the functional form and the BRT analyses the influence of the variables capturing non-linear effects and complex interactions. A challenge of the BRT model is the specification of the hyper-parameters which include the number of trees, the learning rate, and the tree complexity. We used n-fold cross validation to determine the optimum number of trees, and followed the recommendations of \cite{elith_title:_2008} to optimise the remaining parameters to produce an accurate model and minimise risk of over-fitting. For this model we used a tree complexity of 5, and a learning rate of 0.01, with 4100 trees fitted. We implemented the BRT using the 'gbm' and 'dismo' packages in the R programming language. 

A probit regression was carried out for comparison with the BRT, as this is a commonly used model for studies on energy transition concerned with a binary outcome. Assuming that the individual's decision to switch from a biomass stove to a non-biomass stove is based on a latent variable which represents some measure of utility, then this variable can be defined as a linear function of the independent variables, as shown in equation \ref{probit} where $X_{i}$ is a vector of all the independent variables for an individual household, $\beta$ is a vector of coefficients, and $u_{i}$ captures the uncertainty.
\begin{equation}
\label{probit}
   \begin{split}
     y^*_i  = X_{i}\beta + u_{i}
\end{split}     
\end{equation}

The binary outcome we are interested in with these models is not unlike the binary outcomes in medical models assessing patient outcomes (although in our study the outcome is a switch from biomass to non-biomass or not, instead of life and death), and in both cases there is a need for the models to not only perform well on average but also to perform well in distinguishing borderline cases. In the field of medicine when assessing models for patient outcomes it is good practice to report the calibration and discriminatory ability of the model \citep{steyerberg_assessing_2010}. The Brier Score is an overall performance measure of calibration and discrimination for binary outcomes whose scoring rule is shown in equation \ref{Brier} where \textit{N} is the number of instances, \textit{f} is the outcome from the model, and \textit{o} is the actual outcome. The concordance statistic \textit{c}, identical to the area under the Receiver Operating Characteristic (ROC) curve for binary outcomes offers a measure of how well the model distinguishes outcomes. Both of these measures were calculated for each model using base packages in R.

\begin{equation}
   \begin{split}
     Brier Score = \frac{1}{N} \displaystyle\sum_{i=1}^{N}(f_{i} - o_{i})^2
\end{split} 
\label{Brier}
\end{equation}

For the second stage of the analysis hierarchical clustering was used. This is an unsupervised machine learning method that can be used to identify subsets within a dataset that have similar characteristics based on the connectivity between data points. A benefit of hierarchical clustering algorithms for such descriptive analysis is that the iterative process produces a clear tree like structure of clusters which offers a more intuitive view of the clustering process and easier analysis of results, although the iterative nature of the algorithm makes it inefficient for extremely large datasets \citep{kassambra_practical_2017}. We used an agglomerative hierarchical clustering algorithm and with the gower distance measure for categorical variables as it produced a clear and distinct cluster structure. All analysis was performed in R using base packages, as well as the 'dendextend' and 'fpc' packages.

\section{Predictive Modelling Results}

\subsection{Boosted regression tree model}
\begin{figure}[h!]
\centering
\includegraphics[width=\textwidth]{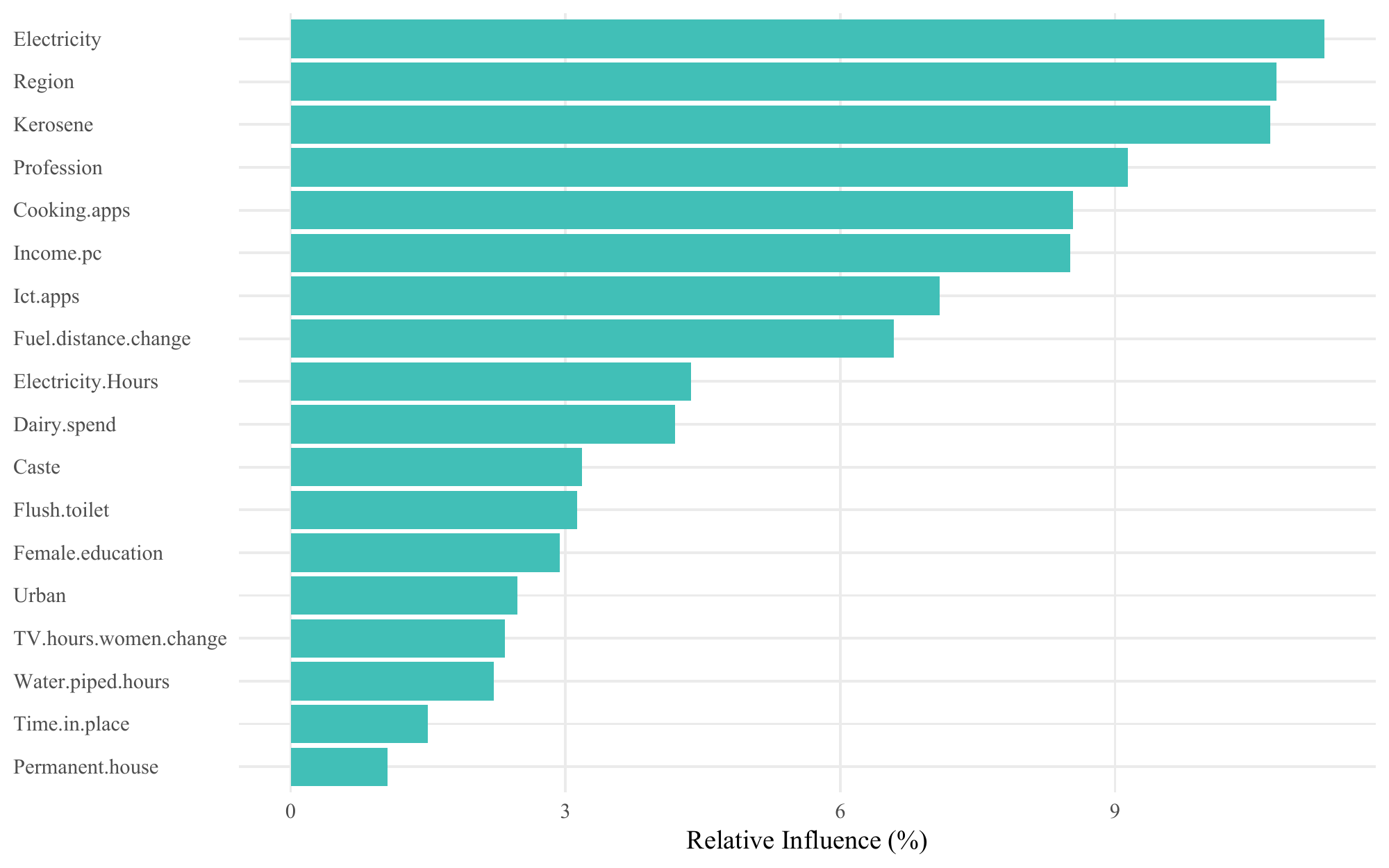}
\caption{Relative influence of variables in BRT Model}
\label{GBM_Marginal_1}
\end{figure}

\begin{figure}[h!]
\centering
\includegraphics[width=\textwidth]{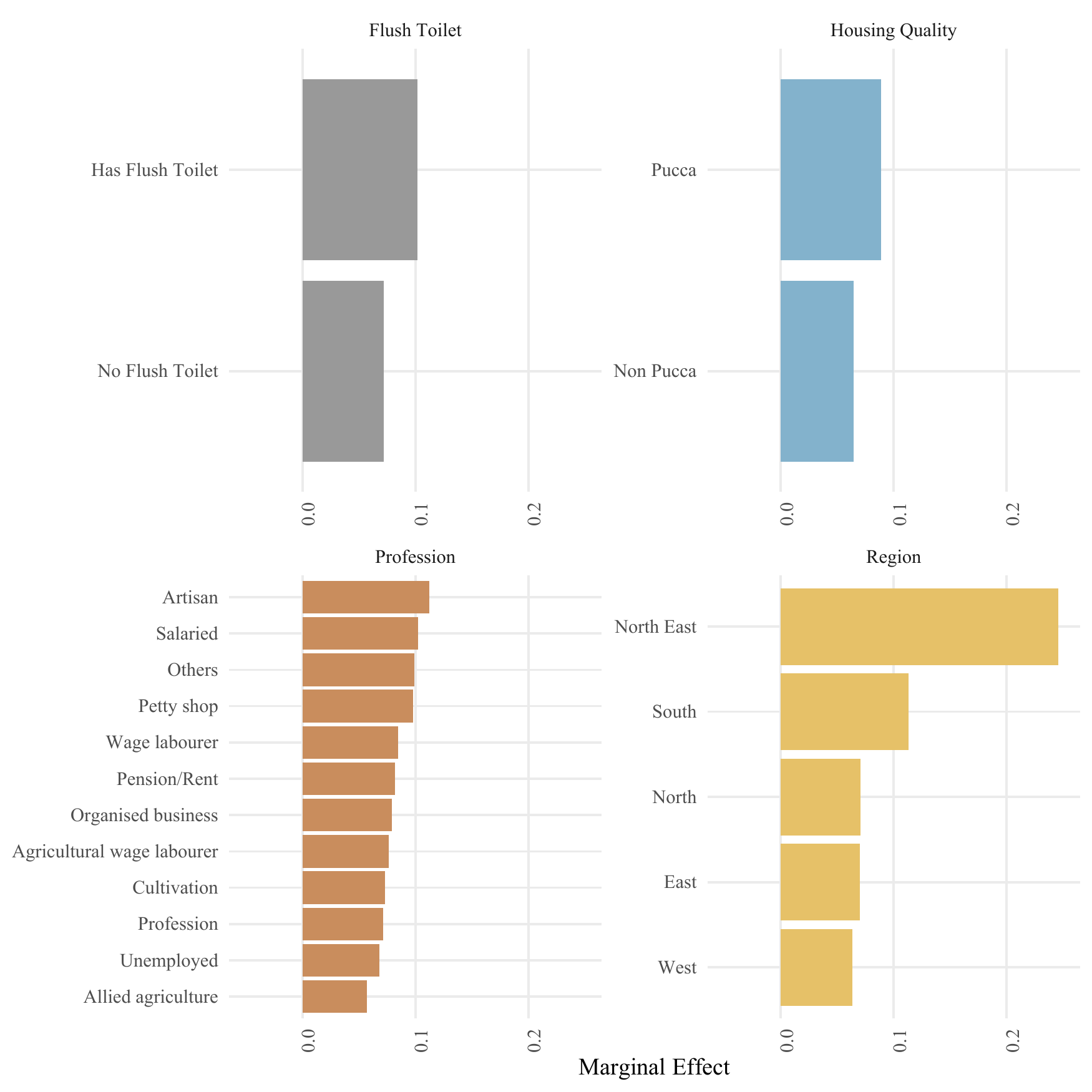}
\caption{Marginal effect of constant effect independent variables on probability of a household switching from Biomass to LPG}
\label{GBM_Marginal_Constant}
\end{figure}

\begin{figure}[h!]
	\centering
	\includegraphics[width=\textwidth]{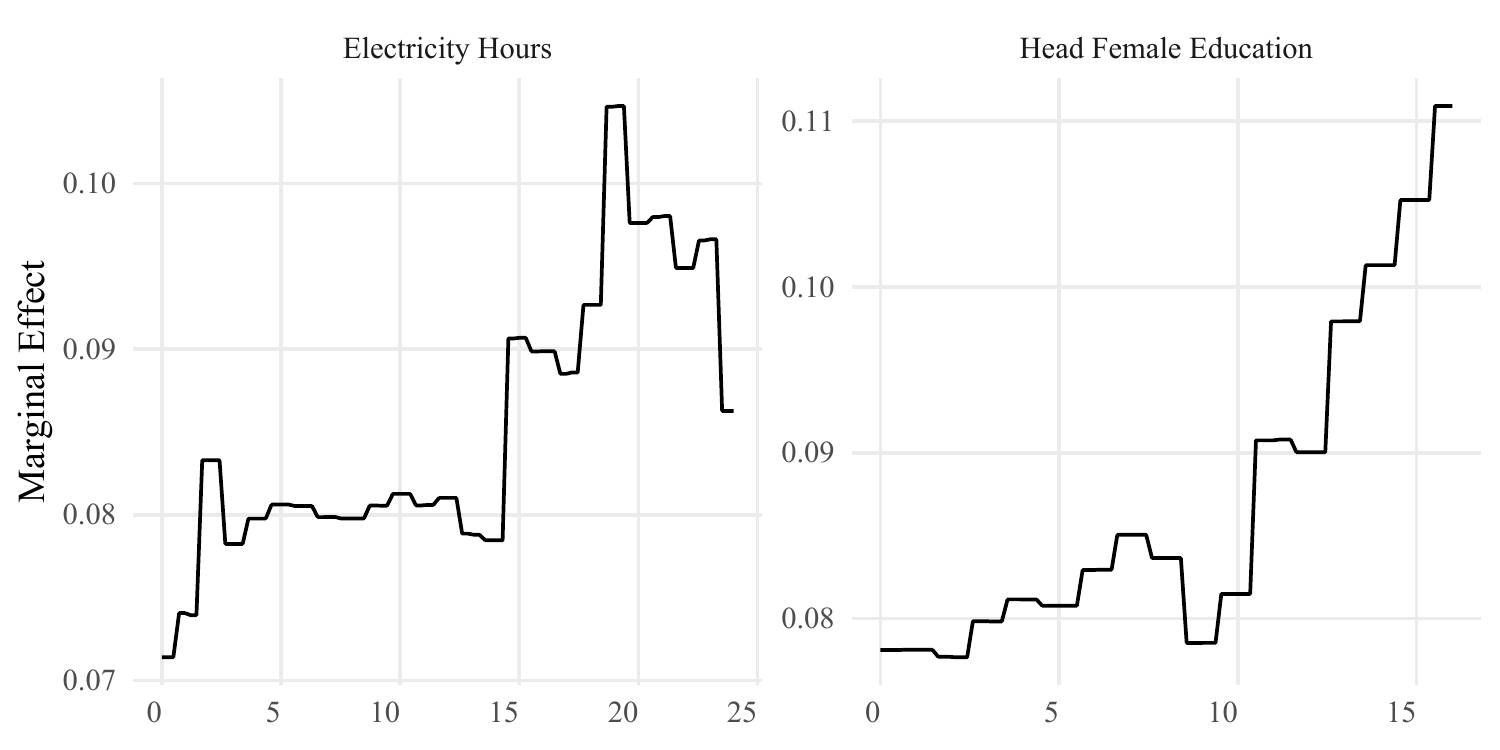}
	\caption{Marginal effect of threshold response independent variables on probability of a household switching from Biomass to LPG}
	\label{GBM_Marginal_Threshold}
\end{figure}

\begin{figure}[h!]
	\centering
	\includegraphics[width=\textwidth]{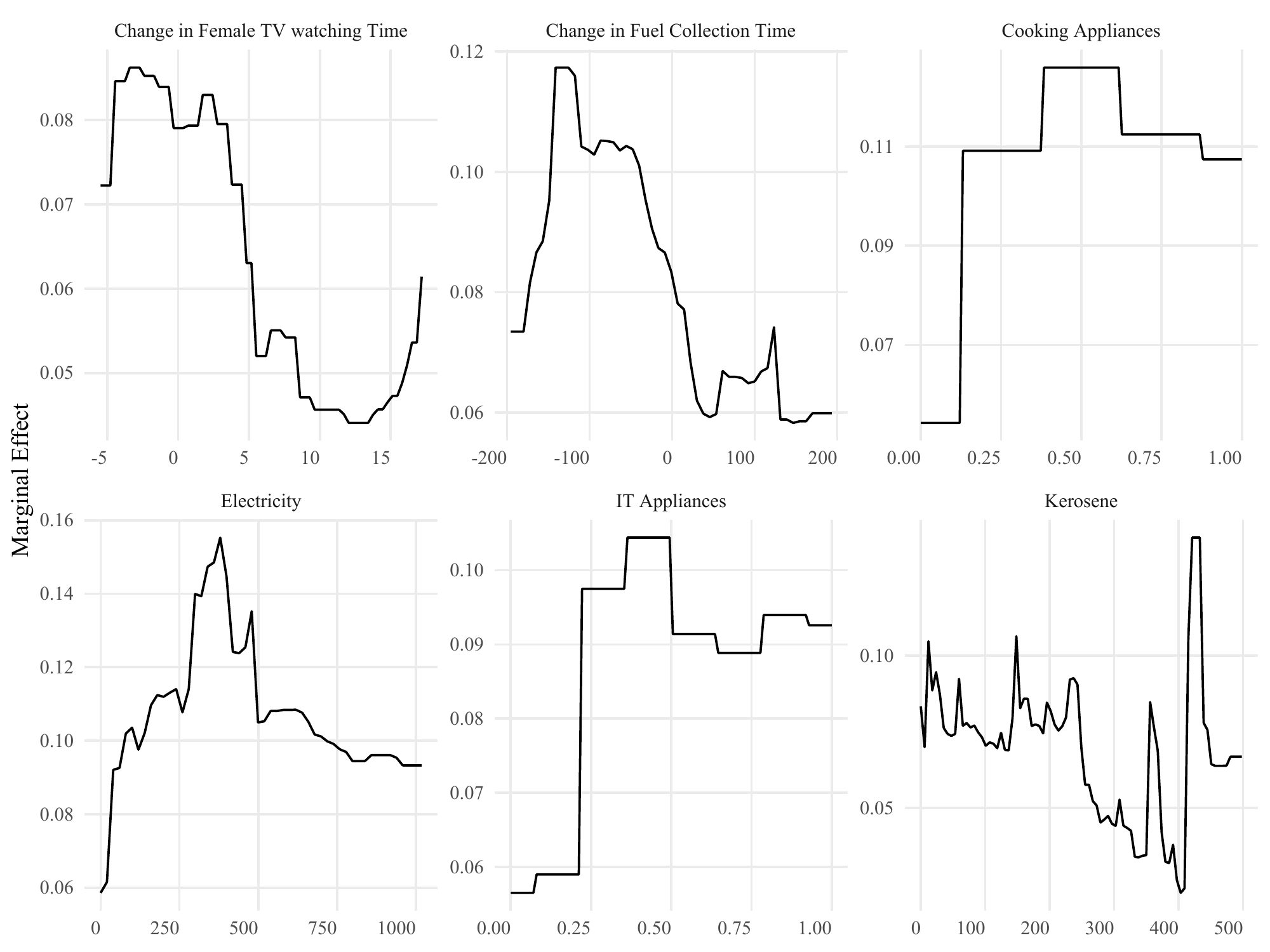}
	\caption{Marginal effect of multiple threshold response independent variables on probability of a household switching from Biomass to LPG}
	\label{GBM_Marginal_Regimes}
\end{figure}

From the BRT analysis we obtain both the relative importance of variables shown in figure \ref{GBM_Marginal_1} and the marginal effects of the independent variables shown in figures \ref{GBM_Marginal_Constant}, \ref{GBM_Marginal_Threshold}, \ref{GBM_Marginal_Regimes}. Figure \ref{GBM_Marginal_1} shows all independent variables were found to have non-zero relative influence ranging from 1-12\%. Use of kerosene and electricity both have an influence of around 11\%, while cooking equipment ownership shows an 8.5\% influence, and IT appliance ownership a 7.1\% influence. The region a household is in has a 10.7\% influence and the profession of the head of the household has an influence of 9.3\%. Income per capita of the household does have an influence of 8.5\% but the BRT shows it is not the dominant determinant of a household's switch to non-biomass stoves. The marginal effects for each variable shown in figures \ref{GBM_Marginal_Constant}, \ref{GBM_Marginal_Threshold}, and \ref{GBM_Marginal_Regimes} exhibit one of three different types of response: either a constant response (for categorical variables), a threshold response, or a multiple threshold (multiple regime) response. 

The constant marginal effects observed for categorical variables shows that these variables will be key determinants of modern stove switching for only some households - for example region is one of the more relatively influential variables, with North-Eastern states being associated with a markedly higher probability of switching stove, while households in the South have a slightly higher chance of switching than households in the East, North and West where region is a determinant of minor influence. This difference could be the result of local policy or climate differences; for example the southern states are typically wealthier relative to the national average, and southern states such as Tamil Nadu and Karnatka have led development in renewable energy infrastructure in India \citep{schmid_development_2012}. North Eastern states have lower incomes and with historically lower access to infrastructure \citep{ghosh_role_1998}, the geography of this region also results in greater local availability and dependency on biomass fuel compared to other regions \citep{bhatt_fuelwood_2016}. LPG distribution infrastructure development under 'Vitrak Yojana' between 2005 and 2011 benefited many poorly serviced settlements in North Eastern states. In the work of \cite{sankhyayan_availability_2019} a significant relationship between region and LPG use was not found, however the coefficients from their model are compatible with the marginal effects from our analysis.

The profession of the head of the household was also found to be of greater relative influence, although the marginal effects were only significant for some professions as shown in figure \ref{GBM_Marginal_Constant}. Those in skilled trades, artisans, salaried employment, or collecting pensions or rent all had a greater probability of switching, whereas those in agricultural wage labour, and unskilled work were less likely to switch. \cite{kemmler_factors_2007} found that more labour intensive and 'daily wage' type employment was associated with lower electricity use, and \cite{sehjpal_going_2014} found that, in rural India, households whose head was in more formal employment had a greater likelihood of the household transitioning to clean cooking. This may be related to the frequency of payment with the former group of jobs being associated with regular monthly or weekly pay whereas income can be more erratic for the latter group.

A measure of household infrastructure is provided through variables measuring permanent house construction, and availability of flush toilets shown in figure \ref{GBM_Marginal_Constant} and both show a small positive increase in marginal effect on the switch to a non-biomass stove with greater levels of access. \cite{rao_white_2017} similarly found that better dwelling quality had a positive relationship with ownership of refrigerators and TVs, and \cite{ahmad_fuel_2015} showed that access to piped water was associated with clean cooking. Permanent housing, while having the lowest relative influence of the variables in the dataset, did have a positive marginal effect on the switch to a non-biomass stove. These findings suggest that access to public utilities and quality of the household's immediate built environment are important, as \cite{debnath_how_2019} found in their study of rehabilitated slum housing in Mumbai.

Figure \ref{GBM_Marginal_Threshold} shows the marginal effects of variables which exhibit a threshold response, namely hours of electricity supply and years of education of the head female of the household. The marginal effect of hours of electricity supply on switching behaviour shows a constant effect up until 15 hours of electricity supply per day, after which the marginal effect increases with hours of electricity. \cite{rao_white_2017} similarly found that hours of electricity supply had a positive relationship with ownership of refrigerators and TVs, and \cite{ahmad_fuel_2015} showed that access to electricity was associated with clean cooking. The threshold observed at 15 hours could be indicative of the added convenience or reliability of having electricity available for two thirds of the day, encouraging investment in appliances or changing household practices related to cooking.

Education of the head female of the household also displays a threshold response as seen in figure \ref{GBM_Marginal_Threshold}. Households whose head female has  10 or more years of schooling, i.e. completing some level of
secondary or tertiary education, has a greater probability of switching to a 'modern stove'. A recent study by \cite{sharma_transition_2019} found a significant relationship between education and LPG uptake for households in the eastern states of Chattisgarh and Jharkhand, while \cite{ahmad_fuel_2015} found female education to be a significant determinant of non-biomass cooking in non-slum households. In their study, \cite{sankhyayan_availability_2019} found that in urban areas there was a stronger positive association between female literacy and LPG use, especially for households where the female head of the household had more than 9 years of schooling, and they suggest this difference is a result of female literacy not translating into female empowerment as effectively in rural households.

Figure \ref{GBM_Marginal_Regimes} shows the marginal effects of variables with multiple thresholds, or different regimes, where marginal effect follows different trends within given ranges. LPG and biomass fuels are used fairly exclusively for cooking. In contrast electricity and kerosene have a range of different end uses. Use of these fuels can indicate transition to cleaner energy for other household activities which offers an explanation for the high relative influence of these variables. In figure \ref{GBM_Marginal_Regimes} we can see that low levels of electricity consumption are associated with a negative marginal effect on the probability of a household switching but this marginal effect increases to a positive level with increasing electricity consumption up to a level of 500kWh/month. Beyond this electricity has a negligible effect on the probability of switching as households using more electricity than that almost certainly have transitioned to clean cooking, with over 80\% of households using no biomass fuel at all. We similarly see that Kerosene use up to 200 kWh leads to a greater probability of a household switching whereas above that 200 kWh the marginal effect is negative indicating reduced chance of switching. Households using more than 200 kWh of kerosene are likely using it for cooking, hence already have a non-biomass stove. The noisy behaviour between 350 kWh and 500 kWh is likely due to households switching from a biomass stove to a kerosene one, which counts as a 'modern stove' switch in the IHDS. The different marginal effect thresholds show how related energy practices of the household shape the observed energy consumption and how these practices have inter-dependencies, as \cite{bisaga_climb_2018} found in their study.

Appliance ownership can serve as a proxy for energy use by a household as appliances are used to deliver a particular energy service. Figure \ref{GBM_Marginal_Regimes} shows how increasing ownership of IT and cooking appliances increases the probability of a household having switched to a non-biomass stove. \cite{rao_white_2017} found that refrigerator and television ownership was associated with greater LPG use by a household, which suggests clean cooking facilities. Greater appliance ownership could also signal better access to markets or shops, as well as better availability of electricity.  However there are two thresholds, as the marginal effect plataeus for households with average ownership, and drops off at high ownership levels as households with very high levels of appliance ownership are more likely to already use LPG and thus the greatest marginal probability of switching occurs for households with middling levels (40-60\%) of ownership. 

Time spent collecting fuel and watching TV in a household shown in figure \ref{GBM_Marginal_Regimes} offer some quantification of household practices as a measure of time allocation to given practices. A decrease in time spent collecting fuel of up to 130 minutes is associated with a greater probability of a switch to a 'modern stove', and decreases in time spent collecting fuel beyond 130 minutes have a relatively low marginal effect on the chance of a household transitioning. An increase up to 50 minutes is associated with a decreasing probability of switching and increases in fuel collection time above 50 minutes see the lowest probability of switching. Similarly the change in number of hours spent watching TV by the adult women of the household has a small positive association for small decreases and increases, but larger increases beyond 5 hours of TV viewing are associated with a lower probability of a household stove switching. The marginal effect of changes in energy practices surrounding energy use and clean cooking transitions are characterised by multiple thresholds. Additionally the marginal effects of these two variables quantitatively shows that there is a change in the time allocated to energy related practices in a household that switches stove. This is important as it implies that characteristics of the stove and its usage have an impact on the practices of a household. \cite{debnath_how_2019} found that characteristics of household appliances in Mumbai slums had a significant effect on the practices of the household.

\subsection{Probit model}
The coefficients of the probit regression model are shown in table \ref{Logit_results}. The same subset of 18 variables used in the BRT model were included in this model, although several profession categories encompassing smaller proportions of the population as well as caste were not found to have a significant effect (at p$<$0.1) on fuel switching, and are not included in the table. A key difference between the outputs of the probit and BRT models is that while the BRT provides relative importance and marginal effect plots, the probit model provides coefficients, standard errors, and confidence intervals denoted by statistical significance levels which can make the process of evaluating the model more straightforward. Comparing the coefficients in table \ref{Logit_results} with the relative importance and marginal effect plots from the BRT model in figures \ref{GBM_Marginal_Constant}, \ref{GBM_Marginal_Threshold} and \ref{GBM_Marginal_Regimes} we can see that many of the coefficients and marginal effects for many of the categorical variables such as region, permanent housing, profession, and flush toilet availability show compatibility with respect to influence on stove switching.

\begin{table}[htbp] \centering 
  \caption{} 
  \label{Logit_results} 
\begin{tabular}{lcc} 
\\[-5ex]\hline 
\hline \\[-4ex] 
 & \multicolumn{2}{c}{\textit{Dependent variable}} \\ 
\cline{2-3} 
\\[-4ex] Independent Variable & Coefficient & Standard Error \\ 
\hline \\[-4ex] 
 Region: North & $-$0.008 & (0.046) \\ 
 Region: North East & 1.078$^{***}$ & (0.089) \\  
 Region: South & 0.494$^{***}$ & (0.045) \\ 
 Region: West &  $-$0.091$^{**}$ & (0.045) \\ 
 Income per capita & $-$0.000006$^{*}$ & (0.000003) \\ 
 Time in place & 0.003$^{***}$ & (0.001) \\ 
 Female education & 0.003 & (0.004) \\ 
 Profession: Agricultural wage labourer & 0.811$^{*}$ & (0.478) \\  
 Profession: Artisan/Skilled & 1.036$^{**}$ & (0.483) \\ 
 Profession: Pension/Rent & 0.847$^{*}$ & (0.477) \\ 
 Profession: Petty shop & 1.038$^{**}$ & (0.477) \\
 Profession: Salaried &  0.949$^{**}$ & (0.477) \\ 
 Profession: Wage labourer & 0.958$^{**}$ & (0.478) \\ 
 Pucca house & 0.352$^{***}$ & (0.037) \\ 
 Flush toilet & 0.266$^{***}$ & (0.035) \\ 
 Water piped hours & 0.007$^{**}$ & (0.003) \\
 Dairy spend & -0.00004 & (0.0001) \\ 
 Electricity hours & 0.010$^{***}$ & (0.002) \\  
 Electricity & 0.00001 & (0.0001) \\ 
 Kerosene & 0.001$^{**}$ & (0.0004) \\ 
 Fuel distance change & -0.002$^{***}$ & (0.0004) \\ 
 Cooking appliance ownership & 0.311$^{***}$ & (0.085) \\ 
 IT appliance ownership & 0.916$^{***}$ & (0.111) \\ 
 TV hours women change & -0.010 & (0.008) \\ 
 Constant & -2.817$^{***}$ & (0.540) \\ 
\hline \\[-4ex] 
Pseudo R$^{2}$ & 0.115 & \\ 

\hline 
\hline \\[-4ex] 
\textit{Note:}  & \multicolumn{2}{r}{$^{*}$p$<$0.1; $^{**}$p$<$0.05; $^{***}$p$<$0.01} \\ 
\end{tabular} 
\end{table}

However there are some key differences between the outputs, particularly those which have a non linear effect in the BRT model. For example, while the BRT identified the use of complimentary fuels as being significant, the probit regression does not find any significant effect. If we look at the marginal effect plots for electricity use in figure \ref{GBM_Marginal_Regimes} we can see that the marginal effects vary with the level of respective fuel use. This non-linear relationship cannot be captured by the probit regression. Conversely while the probit regression correctly identifies significant effects for variables such as cooking and IT appliance ownership, distance travelled for fuel, and hours of electricity supply, it does not capture the threshold identified by the BRT beyond which the marginal effects of these variables are reduced or negligible.

\subsection{Comparison of predictive performance of BRT and Probit Models}
Using the test subset of the dataset as inputs to each of the two models, predictions of whether a household would switch to a non-biomass 'modern stove' or not were calculated and compared to the actual stove switching outcome in the dataset. Table \ref{Pred_perform} shows the classification tallies of each model as well as three measures of predictive performance: the percentage of correctly classified households (a higher score indicates better predictive ability); the AUC score indicating discriminative ability of the model (a higher score indicates better predictive ability); and the Brier score which is an indication of both calibration and discriminative ability of the model (a lower score indicates better predictive ability).

\begin{table}[htbp] \centering 
  \caption{Results of indicators for comparison of predictive performance of BRT and Probit Model} 
  \label{Pred_perform} 
\begin{tabular}{@{\extracolsep{5pt}}lcc} 
\\[-5ex]\hline 
\hline \\[-4ex] 
 & \multicolumn{2}{c}{\textit{Model}} \\ 
\cline{2-3} 
\\[-4ex] & BRT Model & Probit Model  \\ 
\hline \\[-4ex] 
 Correct classification & 84.9\% & 83.5\% \\ 
 AUC & 0.823 & 0.731 \\ 
 Brier Score & 0.108 & 0.126 \\ 
\hline
 True positive & 214 & 103 \\ 
 False negative & 1138 & 1249 \\
 False positive & 89 & 116 \\ 
 True negative & 6866 & 6839 \\ 
\hline
\hline \\[-4ex] 
\textit{Note:}  & \multicolumn{2}{r}{test subset of 7922 households} \\ 
\end{tabular} 
\end{table} 

\begin{figure}[h!]
	\centering
	\includegraphics[width=\textwidth]{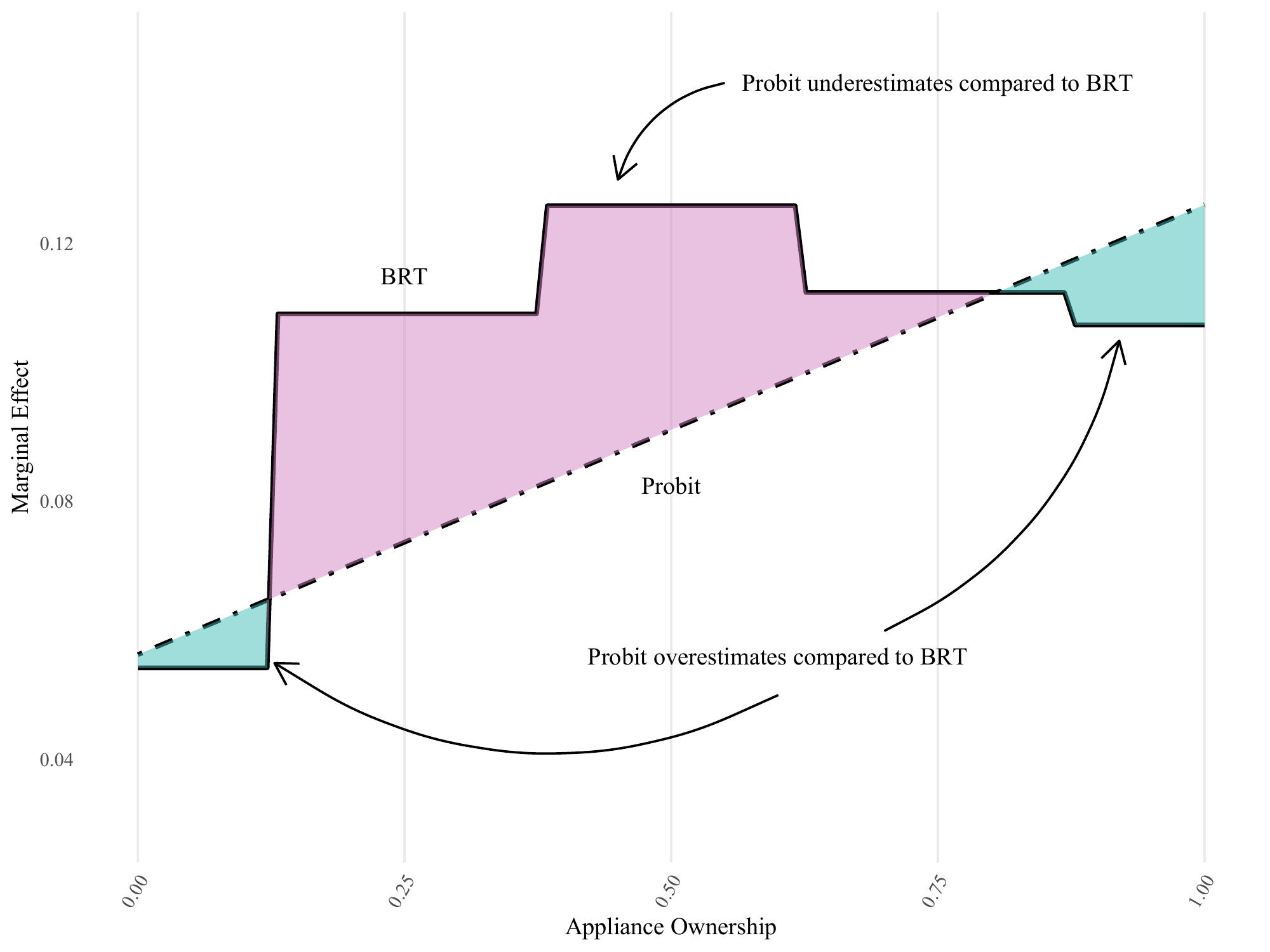}
	\caption{Comparison of Marginal Effect of Cooking Appliance Ownership from Probit and BRT Models}
	\label{GBM_vs_Probit}
\end{figure}

The BRT model outperforms the probit model on all three measures particularly on its discriminatory ability, although the results are comparable. This is a reflection on the ability of the tree-based ensemble method to model non-linear effects. Indeed many of the independent variables had non-linear marginal effects. Thresholds for non-zero effects are a reflection of the non-linear nature of practices and decision-making concerning household energy use. Figure \ref{GBM_vs_Probit} demonstrates this difference between the probit and BRT model using the example of cooking appliance ownership. As shown both models follow the same positive trend with greater appliance ownership and have similar marginal effects at the mean. However, for specific households the probit either under- or overestimates the effect of appliance ownership compared to the BRT model. The probit regression offers the benefit of simplicity, which can make communicating results to a non-technical audience straightforward. Additionally the assessment of compatibility of results via statistical significance can help validate and compare results. However, the outputs of the BRT offer a visual and intuitive way of conveying the variation in marginal effect and the existence of threshold levels.

While our measures of performance provide a metric for the calibration and discriminatory ability of each model, the rates of true and false positives and negatives for each model shown in the bottom half of table \ref{Pred_perform} point to a problem of such models. For both the probit  and BRT models we find that the number of false negatives, that is the households that the model predicted would not switch but did in reality switch, accounts for over 84\% of switching households in the BRT model and 92\% of such households under the probit model. This suggests that while these models are good at predicting households that did not switch (true negatives compared with false positives), they perform poorly at predicting households that do transition. Households that transition against the expectation of the model point to the existence of alternative transition pathways not captured by either model, defined by characteristics that individually would ordinarily not be drivers of transition, but when present in specific combinations can allow households to overcome other barriers.

\section{Descriptive Modelling Results}
\begin{figure}[htbp]
\centering
\includegraphics[width=\textwidth]{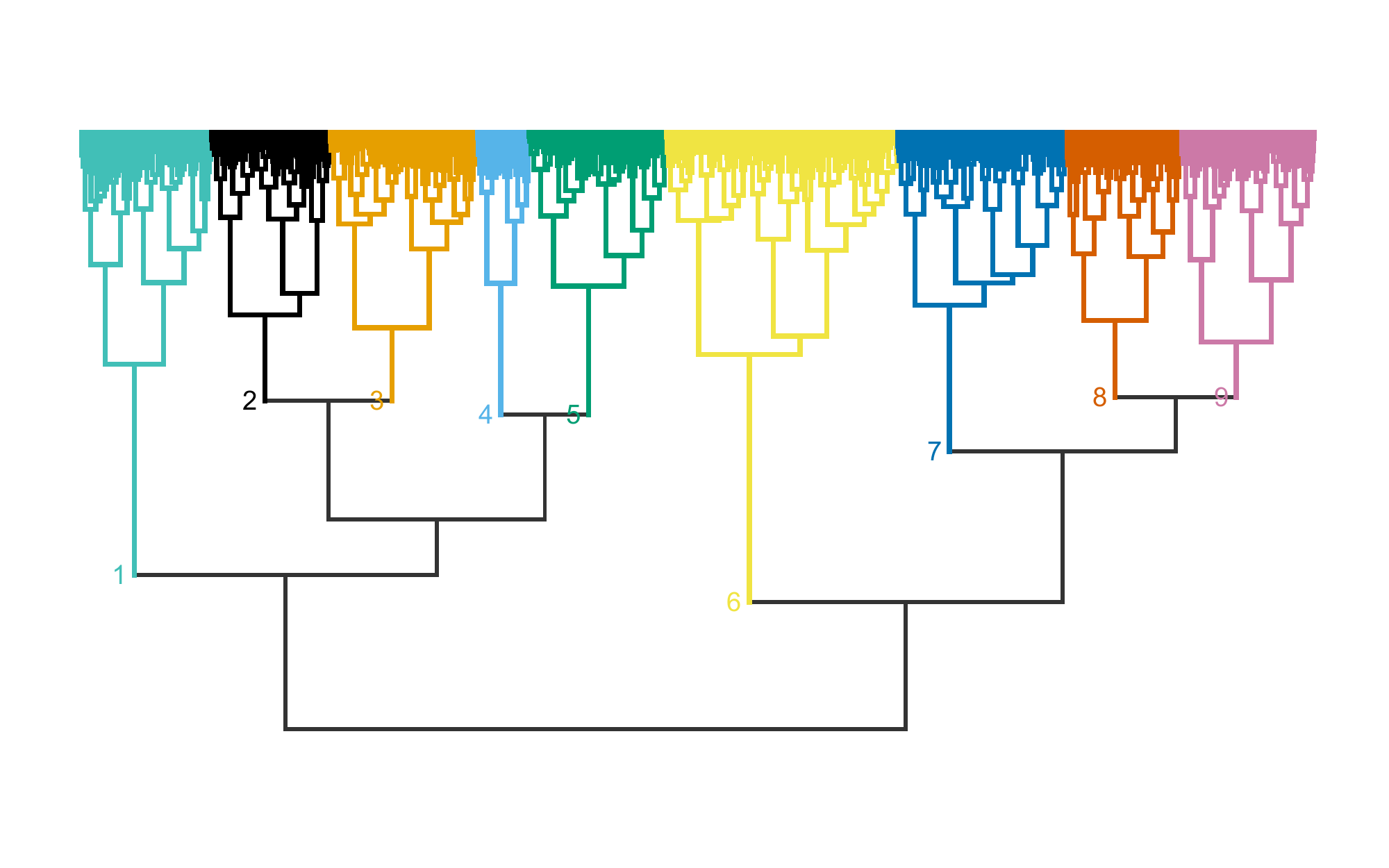}
\caption{Dendrogram of Hierarchical Clustering with IHDS Biomass to LPG switching households}
\label{Dendrogram}
\end{figure}

Using the variables shown in table \ref{Table of Variables} a divisive hierarchical clustering analysis was conducted on the subset of households that did switch their main stove from solid fuel biomass stoves to a clean non-biomass stove between 2004-5 and 2011-12. The clustering analysis identifying nine distinct clusters of households all of which had transitioned away from primarily using a biomass stove but with different combinations of defining characteristics. The resulting dendrogram is shown in figure \ref{Dendrogram}, and the mean characteristics of each cluster are shown in table \ref{Cluster_table}.
\begin{sidewaystable}
\caption{Mean characteristics of clean cooking transition clusters} 
 \label{Cluster_table}
    \centering
    \begin{tabular}{rlllllllll}
  \hline
 & 1 & 2 & 3 & 4 & 5 & 6 & 7 & 8 & 9 \\ 
  \hline
  Region (most represented) & North & South & South & North & North & North East & East & West & South \\ 
  Income & 55058.20 & 33054.90 & 27915.67 & 46401.94 & 29561.38 & 48916.25 & 40480.51 & 35244.72 & 37380.36 \\ 
  Caste & Fwd/Gen & OBC & OBC & OBC & OBC & Fwd/Gen & Fwd/Gen & OBC & OBC \\ 
  Time in Place & 85.18 & 82.87 & 79.91 & 64.58 & 70.67 & 75.81 & 63.07 & 84.60 & 71.86 \\ 
  Urban (\%) & 0.01 & 0.02 & 0.38 & 0.99 & 0.96 & 0.54 & 0.98 & 0.01 & 0.48 \\ 
  Female Education &  8.52 &  5.89 &  5.57 &  7.96 &  5.98 & 10.05 &  8.46 &  6.92 &  8.51 \\ 
  Pucca House & 0.98 & 1.00 & 0.00 & 0.99 & 0.96 & 0.78 & 1.00 & 0.94 & 0.98 \\ 
  Flush Toilet & 0.99 & 0.02 & 0.41 & 1.00 & 0.06 & 0.79 & 0.57 & 0.44 & 0.99 \\ 
  Water Piped Hours & 2.53 & 1.70 & 1.98 & 2.22 & 4.91 & 0.35 & 3.04 & 1.86 & 3.23 \\ 
  Monthly spend on dairy & 461.24 & 220.90 & 162.93 & 301.77 & 194.66 & 355.35 & 186.90 & 185.29 & 154.35 \\ 
  Electricity Hours & 15.14 & 12.55 & 12.96 & 15.79 & 15.56 &  6.67 & 19.61 & 16.59 & 17.98 \\ 
  Electricity & 166.48 &  74.36 &  80.94 & 228.55 & 130.85 &  96.90 & 167.94 &  95.86 & 110.81 \\ 
  LPG & 172.18 & 136.10 & 132.90 & 181.68 & 141.11 & 215.12 & 145.49 & 128.91 & 130.25 \\ 
  Biomass & 466.67 & 522.18 & 423.21 & 159.73 & 327.45 & 360.39 & 131.21 & 598.30 & 380.47 \\ 
  Kerosene & 18.00 & 21.64 & 31.77 & 22.61 & 29.11 & 47.52 & 32.84 & 34.54 & 16.43 \\ 
  Change in Fuel Distance &  -6.46 &  -9.67 &  -9.78 &  -2.30 &  -4.32 &   9.86 &  -4.50 & -21.76 &  -5.06 \\ 
  Cooking Appliances & 0.54 & 0.36 & 0.32 & 0.50 & 0.37 & 0.43 & 0.46 & 0.41 & 0.52 \\ 
  IT Appliances & 0.45 & 0.38 & 0.34 & 0.42 & 0.38 & 0.43 & 0.43 & 0.37 & 0.46 \\ 
  Female TV Viewing Hours & 0.91 & 0.85 & 0.85 & 0.63 & 0.42 & 1.55 & 0.26 & 0.98 & 0.21 \\ 
   \hline
   Correct BRT Prediction & 12.8\% & 8.1\% & 10.0\% & 4.7\% & 13.7\% & 76.6\% & 3.6\% & 12.6\% & 28.2\% \\
   Correct Probit Prediction & 0.6\% & 0.6\% & 0.0\% & 1.0\% & 1.6\% & 64.1\% & 0.0\% & 0.0\% & 24.2`\% \\
   \hline
   
\end{tabular}
\end{sidewaystable}
\begin{figure}[htbp]
\begin{subfigure}{0.45\textwidth}
    \includegraphics[width=\linewidth]{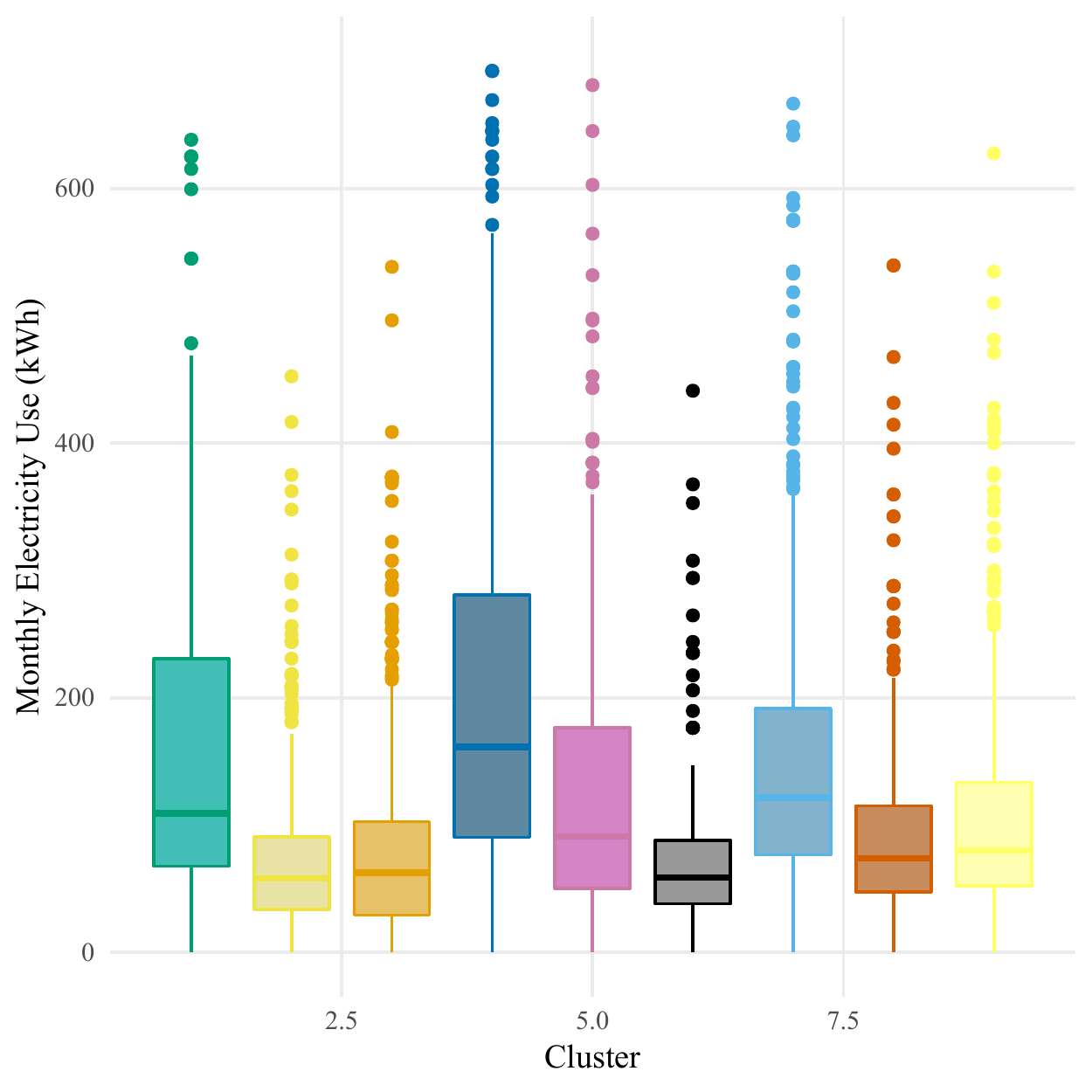}
    \caption{Electricity Use}
    \end{subfigure}
    \hspace{\fill} 
    \begin{subfigure}{0.45\textwidth}
    \includegraphics[width=\linewidth]{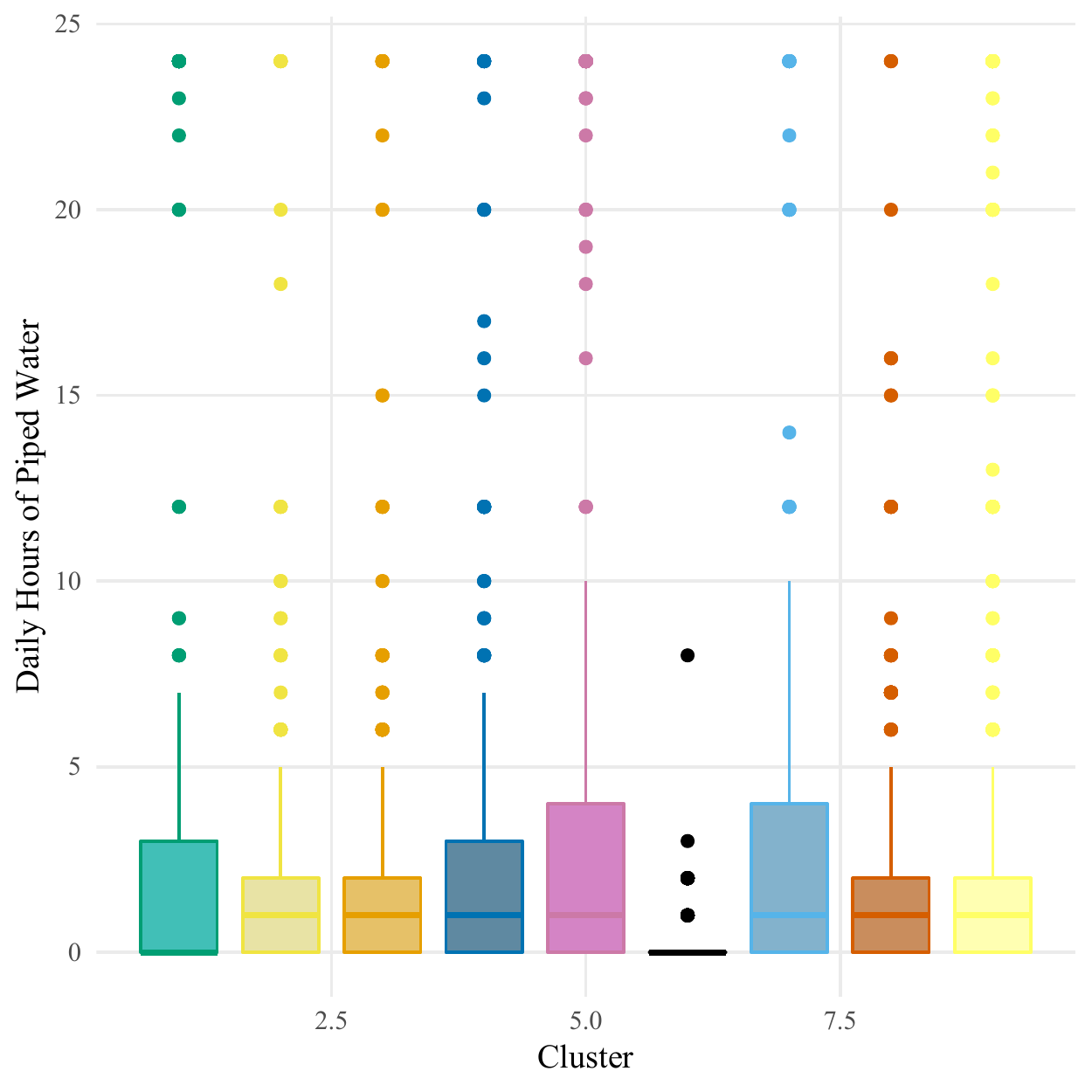}
    \caption{Piped Water Supply}
    \end{subfigure}


    \begin{subfigure}{\textwidth}
    \includegraphics[width=0.95\linewidth]{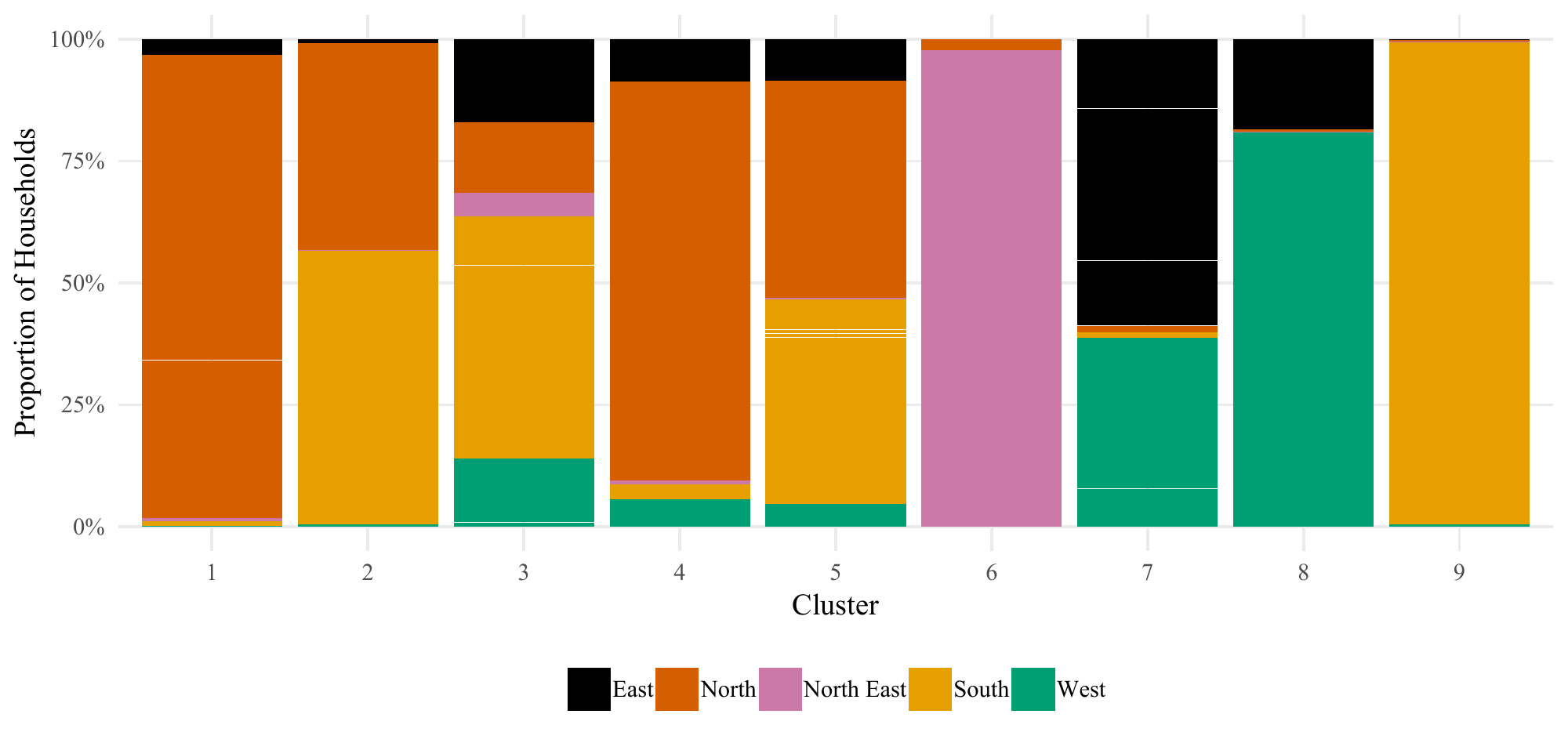}
    \caption{Region}
    \end{subfigure}
    \hspace{\fill} 
    \begin{subfigure}{\textwidth}
    \includegraphics[width=0.95\linewidth]{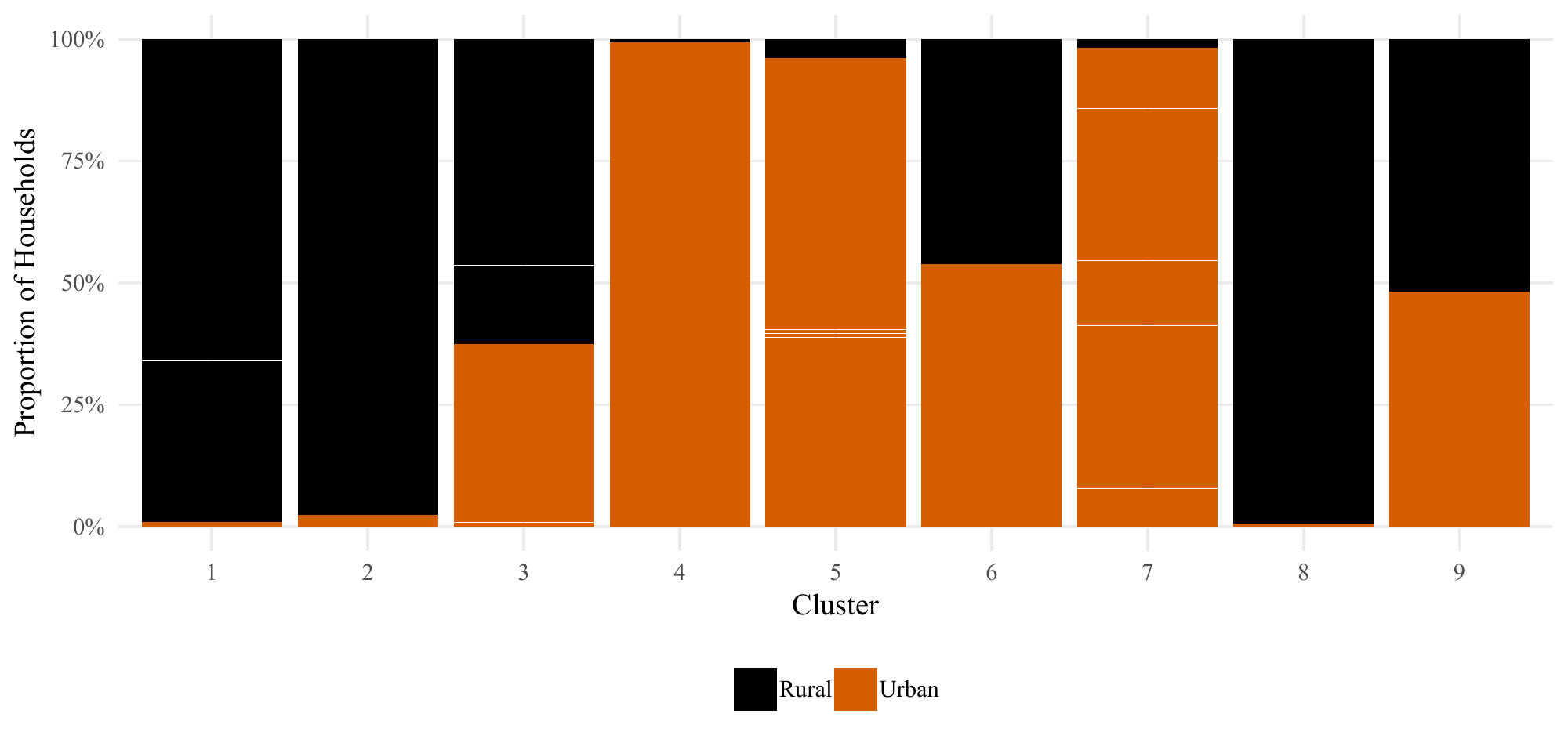}
    \caption{Rural - Urban Split}
    \end{subfigure}
\caption{Key explanatory variables by cluster for households that have switched from Biomass to LPG} \label{fig:4pics}
\end{figure}

The diversity of characteristics between clusters is notable as it suggests that there is no single combination of determinants that results in a transition to clean cooking fuels, and points to the different and complex transition pathways that \cite{van_der_kroon_energy_2013} discussed. A comparison of clusters 1 and 2 detailed in table \ref{Cluster_table} and shown in figure \ref{fig:4pics} serves to illustrate a rural case of such different transition pathways: households in cluster 1 have a mean income of 55,058 INR, and are nearly all Northern rural households. They have good provision of water and electricity, with near ubiquity of flush toilets, permanent housing, above average appliance ownership and electricity use, as well as above average levels of female education. This group represents households that score highly on most of the key determinants, and a higher proportion of these households were correctly predicted to have switched stove by the BRT model. In contrast households in cluster 2 have a lower mean per capita income of 33,054 INR, lower female education levels, lower prevalence of flush toilets, fewer hours of piped water and electricity access, lower electricity consumption and higher average biomass consumption while having lower appliance ownership. However households in cluster 2 all have permanent housing, have been settled for on average over 80 years and still have better than average availability of electricity and water. This suggests that despite their lower income these households still have access to a better than average level of physical infrastructure, but their high biomass use relative to cluster 1 suggests that there is a higher prevalence of fuel stacking in households of cluster 2. 

The existence of different transition pathways can also be observed between urban clusters 4 and 5. Cluster 4 represents above average income households with a per capita annual income of 46,401 INR, and above average education of the head female of the household, access to flush toilets, hours of electricity, and appliance ownership. Cluster 4 also largely represents northern urban households. Households in cluster 5 are also urban, but have markedly lower mean per capita annual income of 29,561 INR, and low prevalence of flush toilets, lower electricity consumption, lower levels of head female education, appliance ownership, and mean biomass consumption double that of cluster 4. Cluster 5 have a high proportion of households employed in stable jobs, and have equally good availability of water and electricity as those households in cluster 4, as well as being settled in their current neighbourhood for longer and containing more Southern households. These longer established households with steady employment are likely to have stronger communities with good 'social infrastructure', with better relationships and sharing of information between neighbours. A greater proportion of households in cluster 5 were correctly predicted to transition by the BRT model as they score highly on key determinants such as region, profession, and change in fuel distance while not lagging too far behind the mean on other key determinants. These households are likely to have a higher prevalence of fuel stacking as evidenced by the higher mean biomass use, where biomass fuels may offer a back up fuel when LPG is not available, or in months when household income needs to be spent on other priorities.

While our predictive models had good performance in predicting non-switching households they did not perform well identifying households that did switch. It is interesting to note the uneven distribution of correct model predictions across the clusters, that is the rate of true positives in the test subset of the dataset present in each cluster. The probit model fails to predict a significant proportion of transitions in any cluster but 6 and 9 where it correctly predicted 64.1\% and 24.2\% of stove transitions respectively. These are the clusters which score highly in nearly all the determinants and are easy identification targets for the model. The BRT model does correctly identify a low percentage of stove switching in several other clusters but similarly performs best at identifying stove switching households in clusters 6 and 9. 

Access to both physical infrastructure - indicated by variables including housing quality, hours of electricity, piped water availability, and flush toilet availability - and/or social infrastructure - indicated by variables including years since migration, caste, and profession - seem to be important to transition to clean cooking. Clusters other than 6 and 9 do not score highly on nearly all determinants, but score highly on different combinations of determinants related to physical and social infrastructure. All of these combinations facilitated a transition albeit following a different pathway. This analysis sheds light on the defining features of each of these transition pathways, and these can indicate key challenges for policy to address to promote adoption and continued use of clean cooking fuel, and reduce dependency on solid biomass fuels. Table \ref{Table of Pathway Policies} summarises four distinct transition pathways observed across the nine clusters, and likely policy challenges associated with households on each pathway.

LPG consumption in rural and peri-urban households such as those on pathway B in table \ref{Table of Pathway Policies}, as well as average income urban and peri-urban households of pathway D suggests that canisters are being refilled less than once a month (a regular 14.2kg canister has approximately 200kWh worth of LPG, so monthly use below this indicates non-monthly refills). This indicates that LPG is not being used to meet all of a households cooking needs, and this could point to either issues of affordability and cash flow with households unable to afford more frequent refills, or issues of supply and delivery which may particularly affect rural and peri-urban households. \cite{sharma_transition_2019} observed that doorstep delivery of LPG canisters in an area increased LPG usage in Chhattisgarh, and noted that often rural households had to collect their cylinders form the distributors. Our analysis indicates that further policy intervention may be needed to address this issue for rural and peri-urban households.

Table \ref{Table of Pathway Policies} also shows that some households may require less policy intervention to encourage uptake of LPG because they exhibit many shared features with households that already use LPG as described for pathway A. Their good access to infrastructure within an urban setting and stronger financial position means that they may only need a small nudge to switch to LPG. These groups also appear more likely to discontinue biomass use once using LPG, and thus need less policy intervention to transition away from biomass stove use. Similarly some households may face region specific challenges that require additional local intervention, such as north-eastern households of pathway C in table \ref{Table of Pathway Policies}, who despite above average socio-economic circumstances and regular LPG use still are dependant on biomass fuel. This could be in part due to different practices with biomass being used for non-cooking needs, or biomass serving as a more reliable alternative to less reliable supply of cleaner fuels.


\begin{table} \centering 
  \caption{Key pathway features of households that transitioned to LPG and relevant policy issues} 
  \label{Table of Pathway Policies} 
\begin{tabular}{p{0.5cm}p{4.5cm}p{1cm}p{4cm}} 
\\[-5ex]\hline 
\hline \\[-4ex] 
Path & Key Pathway Features & Cluster & Policy Issues \\ 
\hline \\[-4ex] 
A & Urban pucca households, above average income, appliance ownership, electricity use and good infrastructure access. Low residual biomass use & 4,7 & Many households of this group will have already switched, and do not continue use of biomass fuels \\ \hline
B & Rural pucca households with average to above average income, electricity use, and good infrastructure. Above average reduction in fuel collection & 1,2,8 & LPG uptake has not stopped regular biomass use. Infrequent LPG refill suggests supply or cost issue. LPG has delivered notable time savings. \\ \hline
C & North eastern households, above average income and appliance ownership but poor infrastructure and low electricity use. Substantial use of biomass fuel. & 6 & Heavy continued reliance on biomass, perhaps due to regional fuel supply challenges or local practices. \\ \hline
D & Urban or peri-urban households with average income, either in pucca or non-pucca housing but good infrastructure access, and continued biomass or kerosene dependence & 3,5,9 & Biomass and kerosene may be relied on as backup. Possible cash flow or supply challenges persist. \\ \hline
\hline \\[-4ex] 
\end{tabular} 
\end{table} 

An important policy challenge common to several clusters is the continued regular use of biomass fuels amongst a majority of households that have adopted a non-biomass stove. Across the country it appears that rural, and peri-urban households that switch to non-biomass stoves still use considerable amounts of biomass fuel, as do urban households on below average income. For many of these households it may be a way of coping with unreliability in supply or access to LPG, or as a means of reducing consumption of LPG to manage household cash flow. It could also arise due to energy related behaviours which favour biomass use, such as a preference for rice over bread. Crucially, the prevalence of fuel stacking to manage energy services as described by \cite{van_der_kroon_energy_2013}, means it is important to recognise that even once non-biomass stoves are adopted, further intervention to change energy related behaviours or improve reliability of LPG supply may be needed to reduce biomass use by these households. Understanding the energy practices and decisions leading to such fuel stacking behaviours requires an understanding at a household level such as demonstrated by \cite{khalid_homely_2017} in order to enable policy interventions to promote greater uptake of sustained clean cooking among such households.


\section{Limitations and Future Work}

This analysis does face a limitation due to the nature of the IHDS dataset which is representative at the national level. It serves to make some crucial comparisons between regions and states, but we cannot locate where households on a particular transition pathway are beyond their region and rural or urban designation. Differences in the non-income drivers that determine clean cooking transitions and the interaction between physical and social infrastructure and household energy practices all take place at a local scale. Larger sample size surveys at a city scale could be used to identify and characterise the different transition pathways of different groups of households. Additional data on the current fuels used, different energy end uses within a household and time of use, as well as aspirations of households would be invaluable. In addition such detailed surveys could include some qualitative interviews with households discussing their energy practices and decisions to provide context to the data. For example this could provide an understanding of the non-monetary trade-offs considered by households when switching to LPG.

The authors note that promisingly a number of recent studies including by \cite{debnath_how_2019} and \cite{sharma_socio-economic_2019} in this journal have carried out local case studies exploring the influence of non-income drivers on changes in energy practices, appliance ownership, and fuel use. Further work with larger and more widely representative samples of such local data is needed while embracing alternative analytical tools such as ensemble methods and clustering analyses alongside qualitative approaches which can help identify the complex action of non-income factors and identify different pathways to transition.

\section{Conclusions and Policy Implications}

This study has used unsupervised machine learning methods in a two stage analysis using predictive modelling to characterise the non-income determinants of a switch from a biomass to a non-biomass stove by Indian households, and descriptive modelling to identify groups of households which had adopted non-biomass stoves with similar energy transition pathways. Using the panel IHDS dataset with over 32,000 households surveyed in 2004/5 and 2011/12, this study uses ensemble machine learning predictive modelling and descriptive clustering analysis to identify households that are missed by current policy interventions.

North-eastern and southern households had a greater probability of switching from a biomass to non-biomass stove, as did those whose head of household was employed in non-manual labour professions. Several determinants displayed a threshold relationship with stove switching, and were only influential determinants of stove switching beyond a given value - for example availability of electricity above 15 hours a day was associated with increased stove switching. Similarly where the head female of the household had more than 10 years of education a similar increasing probability of stove switching was observed. The influence of other determinants was characterised by multiple thresholds or regimes for example low appliance ownership of both cooking and IT appliances had a plateau of greatest marginal effect for households with ownership between 10 and 50\% with slightly lower probability of fuel switching for households with higher appliance ownership and negligible chance of switching below this range.

Our study found that the BRT model performed better than the probit model in predicting whether households switched, however both models performed relatively poorly in identifying the households that did switch compared to those that did not. The clustering analysis showed that there were nine clearly distinguishable groups of household that had switched. Each cluster is defined by different combinations of key determinants. However nearly all the households correctly identified by the predictive models were grouped in only two of the clusters. The other groups of households represent those typically missed out by predictive models and policies informed by such models. The two stage approach in this study provided additional insight over simple predictive models by determining not only the trends in the data but also the latent groups of households within the sample which followed different cooking transition pathways.

There are several major implications from this study for policy interventions aiming to alleviate energy poverty and promote transition to sustained use of cleaner cooking fuels. Firstly households following different energy transition pathways, pose different challenges for policy to promote clean cooking and reduce reliance on biomass. Many rural and peri-urban households may face limitations in supply of LPG, and interventions to address reliability and convenience of supply may promote more sustained uptake of LPG.

Regional infrastructure, climatic, and cultural factors may also constrain adoption of clean cooking fuels, such as in the North East of India, and additional local policies to improve infrastructure or address local energy related behaviours surrounding biomass are needed to promote clean cooking in the region. This supports a conclusion of \cite{kebede_can_2002} that local variations must be factored into the design and tailoring of policies.

Crucially amongst households that do adopt non-biomass stoves, those in rural and peri-urban areas, as well as urban households on average incomes, are likely to continue using considerable amounts of biomass. Policy interventions promoting uptake of clean cooking for such households must pair LPG promotion with interventions to change behaviours to reduce biomass in order to deliver on the public health aims of clean cooking.


\section{Acknowledgements}
The authors are grateful for EPSRC support through the CDT in Future Infrastructure and Built Environment (EP/L016095/1) and to Indian Institute of Human Settlements, India. AP Neto-Bradley is supported by the The Leathersellers' Company and Fitzwilliam College, Cambridge.

\section{Data Availability}
Datasets related to this article can be found at the following online repositories: IHDS-I dataset - http://doi.org/10.3886/ICPSR22626.v8, and  the IHDS-II dataset - http://doi.org/10.3886/ICPSR36151.v2 an open-source online data repository hosted by the Inter-university Consortium for Political and Social Research (ICPSR) (Desai et al. 2010)

\section*{References}

\bibliography{Energy_Policy_IHDS_Paper}
\end{document}